\documentclass[aps,prl,twocolumn,superscriptaddress]{revtex4}

\usepackage{graphicx,color,ulem}

\begin{document}

\title{Doping evoluton of antiferromagnetic order and structural distortion
in LaFeAsO$_{1-x}$F$_x$}
\author{Q. Huang}
\affiliation{NIST Center for Neutron Research, National Institute of Standards and
Technology,Gaithersburg, MD 20899-6102, USA }
\author{Jun Zhao}
\affiliation{Department of Physics and Astronomy, The University of Tennessee, Knoxville,
Tennessee 37996-1200, USA }
\author{J. W. Lynn}
\affiliation{NIST Center for Neutron Research, National Institute of Standards and
Technology,Gaithersburg, MD 20899-6102, USA }
\author{G. F. Chen}
\affiliation{Beijing National Laboratory for Condensed Matter Physics, Institute of
Physics, Chinese Academy of Sciences, P. O. Box 603, Beijing 100080, China }
\author{J. L. Luo}
\affiliation{Beijing National Laboratory for Condensed Matter Physics, Institute of
Physics, Chinese Academy of Sciences, P. O. Box 603, Beijing 100080, China }
\author{N. L. Wang}
\affiliation{Beijing National Laboratory for Condensed Matter Physics, Institute of
Physics, Chinese Academy of Sciences, P. O. Box 603, Beijing 100080, China }
\author{Pengcheng Dai}
\email{daip@ornl.gov}
\affiliation{Department of Physics and Astronomy, The University of Tennessee, Knoxville,
Tennessee 37996-1200, USA }
\affiliation{Neutron Scattering Science Division, Oak Ridge National Laboratory, Oak
Ridge, Tennessee 37831-6393, USA }

\begin{abstract}
We use neutron scattering to study the structural distortion and
antiferromagnetic (AFM) order in LaFeAsO$_{1-x}$F$_{x}$ as the system is doped
with fluorine (F) to induce superconductivity. In the undoped state, LaFeAsO
exhibits a structural distortion, changing the symmetry from tetragonal
(space group $P4/nmm$) to orthorhombic (space group $Cmma$) at 155 K, and
then followed by an AFM order at 137 K. Doping the system with
F gradually decreases the structural distortion temperature, but suppresses
the long range AFM order before the emergence of superconductivity.
Therefore, while superconductivity in these Fe oxypnictides can survive in
either the tetragonal or the orthorhombic crystal structure, it competes
directly with static AFM order.
\end{abstract}

\maketitle




\section{INTRODUCTION}
A determination of the phase diagram in the FeAs-based $R$FeAsO$_{1-x}$F$_{x}$ (where $R=$ La,Nd,Sm,Pr,...) family of high-transition
temperature (high-$T_{c}$) superconductors \cite{kamihara,chen,gfchen,zaren,hhwen} is the first step necessary for a
comprehensive understanding of their electronic properties. The parent
compounds of these FeAs-based materials are nonsuperconducting semimetals.
When cooling down from room temperature, $R$FeAsO first exhibits a
structural phase transition, changing the crystal symmetry from tetragonal (space group $P4/nmm$) 
to orthorhombic (space group $Cmma$), and then orders antiferromagnetically with a spin structure
as shown in Figs. 1a and 1b \cite{cruz,zhao,mcguire,chen1,huang,zhao1,goldman}. While earlier work had shown
that superconductivity induced by F-doping suppresses both the structural
phase transition and static antiferromagnetic (AFM) order \cite{cruz}, how
this process occurs in $R$FeAsO$_{1-x}$F$_{x}$ as a function of F-doping is
still unclear. For example, in a systematic study of\ the F-doping
dependence of the structural and magnetic phase diagram of CeFeAsO$_{1-x}$F$_{x}$, Zhao {\it et al.} \cite{zhao} found that the Fe AFM order disappears before
the appearance of superconductivity. However, it was not clear whether the
orthorhombic structural distortion in the undoped compound is still present
in the underdoped superconducting materials. On the other hand, while
systematic X-ray diffraction experiments on SmFeAsO$_{1-x}$F$_{x}$ reveal
that orthorhombic symmetry is present in the underdoped superconductors,
there are no neutron scattering experiments to directly probe the AFM phase
boundary in these materials \cite{margadonnna}. Finally, recent $\mu $SR,
transport, and M$\mathrm{\ddot{o}}$ssbauer experiments on the phase diagram
of LaFeAsO$_{1-x}$F$_{x}$ suggest a first-order-like phase transition
between the AFM and superconducting phases \cite{luetkens}. Furthermore,
these authors argue that the tetragonal to orthorhombic structural phase
transition is associated with the doping-induced AFM to superconductivity
phase transition, a result clearly inconsistent with Ref. \cite{margadonnna}.

The difficulty in obtaining a comprehensive phase diagram of $R$FeAsO$_{1-x}$
F$_{x}$ stems from the fact that various local probes such as $\mu $SR and M$\rm 
\ddot{o}$ssbauer can detect magnetic long range order but are insensitive to
the structural distortion \cite{luetkens}. On the other hand, X-ray
scattering\ is sensitive to structural distortion\ but does not directly
probe the AFM order. Neutron scattering is capable of detecting both
structural and magnetic order, but requires large sample masses and
therefore has only been done for\ a limited doping range in CeFeAsO$_{1-x}$F$
_{x}$ \cite{zhao}. In this paper, we present a systematic neutron scattering
investigation of LaFeAsO$_{1-x}$F$_{x}$ that complements earlier work on
this system \cite{cruz,mcguire,luetkens}. Our data reveal that the orthorhombic
structural distortion extends beyond the AFM phase and coexists with
superconductivity, whereas there is no evidence of static long range AFM
order coexisting with superconductivity.

\begin{figure}[tbp]
\includegraphics[scale=.7]{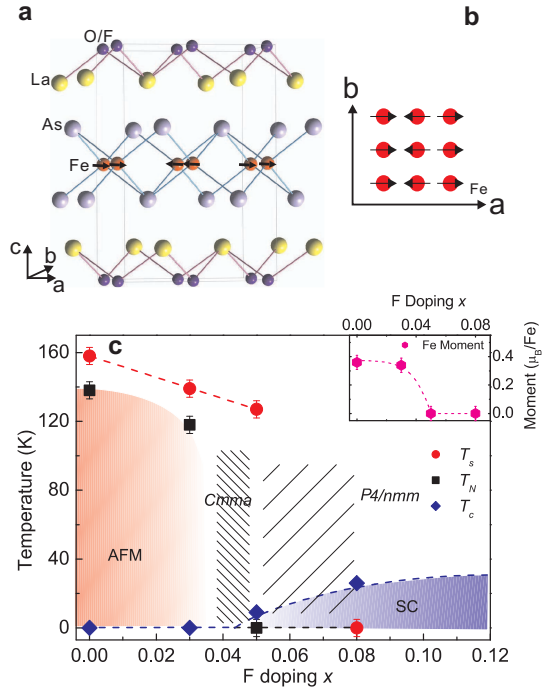}
\caption{(Color online) (a) The Fe spin ordering in the LaFeAsO chemical
unit cell. (b) The Fe magnetic unit cell of LaFeAsO in the Fe-As plane. The
Fe moments lie in the $a$-$b$ plane along the $a$-axis and form an
antiferromagnetic collinear spin structure similar to BaFe$_{2}$As$_{2}$,
SrFe$_{2}$As$_{2}$, and CaFe$_{2}$As$_{2}$ \cite{cruz,huang,zhao,goldman}. (c) The structural and magnetic phase diagram
determined from our neutron measurements on LaFeAsO$_{1-x}$F$_{x}$ with $x=0,0.03,0.05,0.08$. 
The red circles indicate the onset temperature of the 
$P4/nmm$ to $Cmma$ phase transition. The black squares designate the N\'{e}el
temperatures of Fe as determined from neutron measurements in Fig. 3. The
superconducting transition temperatures  $T_{c}$ for $x=0.05,0.08$ are
determined from susceptibility measurements. The AFM to superconducting
phase transition happens between $x=0.03$ and $0.05$. The inset in d) shows
the F doping dependence of the Fe moment as determined from the intensity of
the (1,0,3)$_{M}$ magnetic peak at 4 K. }
\end{figure}

\section{EXPERIMENTAL RESULTS and DISCUSSIONS}

We use neutron diffraction to study the structural and magnetic phase
transitions in polycrystalline samples of LaFeAsO$_{1-x}$F$_{x}$ with fluorine doping $
x=0,0.03,0.05,$ and $0.08$. Our experiments were performed on the BT-1 high
resolution powder diffractometer and BT-7 triple axis spectrometer at the
NIST Center for Neutron Research, Gaithersburg, Maryland. 
The BT-1 diffractometer has a Ge(3,1,1) monochromator and an incident wavelength of 
$\lambda=2.0785$ \AA.  Collimators with horizontal divergences of 
15$^{\prime}$, 20$^{\prime}$, and 7$^{\prime}$ full-width-at-half-maximum (FWHM) 
were used before and after the monochromator, and after the sample, respectively.
The BT-7 has a PG(0,0,2) (pyrolytic graphite) monochromator with
an incident beam wavelength of $\lambda=2.359$ \AA. A PG filter was placed in the incident beam path 
to eliminate $\lambda/2$ \cite{cruz,zhao}.  The collimations are 50$^\prime$ FWHM before the sample and 80$^\prime$ radial collimator between the sample and a position sensitive 
detector that covered an angular range of approximately five degrees.  The polycrystalline 
samples of LaFeAsO$_{1-x}$F$_{x}$ with $
x=0,0.03,0.05,$ and $0.08$ were prepared by the solid state reaction using LaAs, 
Fe$_2$O$_3$, Fe and LaF$_3$ as starting materials, with the detailed 
preparation method described in Ref. \cite{gfchen1}.  We checked the superconducting properties of each LaFeAsO$_{1-x}$F$_{x}$ using a SQUID magnetometer and found that the $x=0,0.03$ samples 
are nonsuperconducting, while the $x=0.05$ and 0.08 samples are 8 K and 26 K superconductors, respectively.
The fluorine-doping levels are nominal, and should be close to the actual electron-doping level at these
concentrations.

\begin{table}[tp]
\caption{Refined crystal structure parameters of LaFeAsO$_{1-x}$F$_x$ with $
x=0$ at 175 K ($R_p =5.24\%$, $w_{Rp} = 6.62\%$, $\protect\chi^2=0.9825$),
and $x=0.08$ at 10 K ($R_p =5.05\%$, $w_{Rp} = 6.6\%$, $\protect\chi%
^2=0.9273 $). Space group: $P4/nmm$. LaFeAsO, $a=4.03007(9)$, $c=8.7368(2)$ 
\AA ; LaFeAsO$_{0.92}$F$_{0.08}$, $a=4.02005(4)$, $c=8.7032(1)$ \AA . }
\label{Table}
\begin{ruledtabular}
\begin{tabular}{cccccccccc}
Atom & site & x & y & {z($x=0$)} & {z($x=0.08$)} &\\[3pt]
\hline&  \\[3pt]
La & $2c$ & ${1}\over {4}$ & ${1}\over {4}$ & 0.1417(3)  & 0.1450(3)  &\\[3pt]
Fe & $2b$ & $3\over 4$     & $1\over 4$     & $1\over 2$ & $1\over 2$ &\\[3pt]
As & $2c$ & $1\over 4$     & $1\over 4$     & 0.6507(4)  & 0.6520(3)  &\\[3pt]
O  & $2a$ & $3\over 4$     & $1\over 4$     & 0          & 0          &\\[3pt]
\end{tabular}
\end{ruledtabular}
\end{table}

Figures 1a and 1b show the Fe spin structure within the FeAs layer as
determined from previous neutron scattering work on $R$FeAsO \cite
{cruz,zhao,mcguire} and (Ba,Sr,Ca)Fe$_{2}$As$_{2}$ \cite{huang,zhao1,goldman}
. Figure 1c summarizes the electronic phase diagram of LaFeAsO$_{1-x}$F$_{x}$
determined from our measurements. Our data are consistent with previous
neutron scattering \cite{cruz} and results from local probes such as $\mu$SR and 
$^{57}$Fe M$\rm \ddot{o}$ssbauer spectroscopy \cite{luetkens}, and
indicates that long range AFM order disappears as a function of doping
before superconductivity is present. \ On the other hand, we find direct
evidence for the orthorhombic structural distortion in the underdoped
superconducting LaFeAsO$_{1-x}$F$_{x}$, indicating that the orthorhombic
lattice distortion extends into the superconductivity dome in LaFeAsO$_{1-x}$
F$_{x}$, similar to that of SmFeAsO$_{1-x}$F$_{x}$ \cite{margadonnna}.

\begin{figure}[tbp]
\includegraphics[scale=1.2]{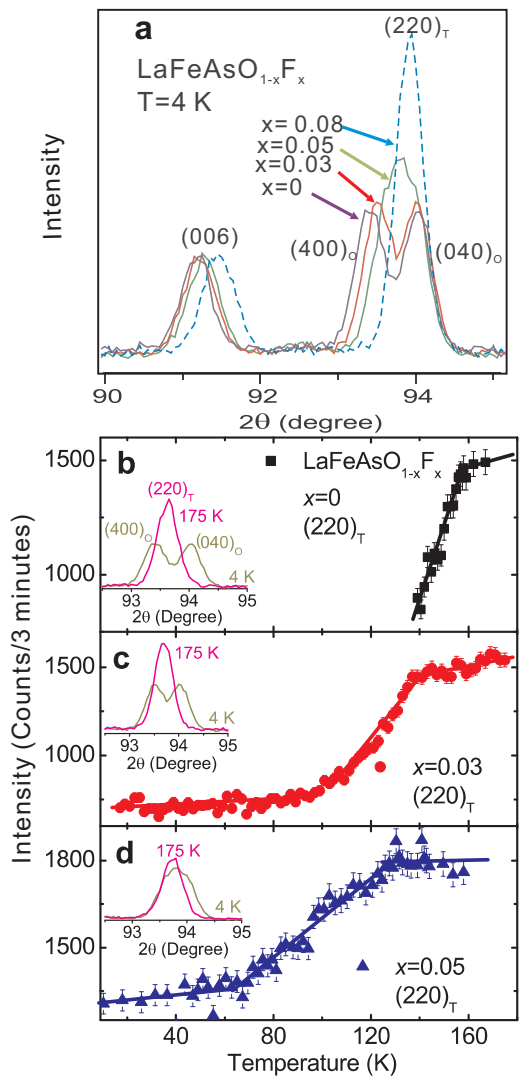}
\caption{(Color online) Dependence of the low-temperature crystal structure
of LaFeAsO$_{1-x}$F$_{x}$ as a function of F-doping $x$. (a) $2\theta $ scans, showing the reduction of the orthorhombic lattice distortion
with increasing F-doping. The $(2,2,0)$ peak for $x=0.05$ is clearly broader
than the resolution. (b-d) Temperature dependence of the $\rm (2,2,0)_{T}$ ($\rm T$
denotes tetragonal) nuclear reflection indicative of a structural phase
transition for various $x$ \cite{cruz}. The temperature of the
tetragonal to orthorhombic lattice distortion reduces with increasing $x$.
The insets show the $\rm (2,2,0)_{T}$ reflection above and below the transition
temperatures. }
\end{figure}

To demonstrate this, we show in Figure 2a a comparison of the
high-resolution BT-1 data for LaFeAsO$_{1-x}$F$_{x}$ with $x=0,0.03,0.05,$
and $0.08$ taken at 4 K. While the parent compound LaFeAsO shows clear
evidence of the orthorhombic lattice distortion as illustrated by the
splitting of the $\rm (4,0,0)_{o}$ and $\rm (0,4,0)_{o}$ peaks, doping F gradually
reduces the splitting of these peaks until they become a single
resolution-limited peak corresponding to tetragonal symmetry for $x=0.08$ 
\cite{cruz}. For $x=0.03$, one can see a clear splitting of the $\rm (4,0,0)_{o}$
and $\rm (0,4,0)_{o}$ peaks. Although a well-resolved splitting is no longer
observable in the $x=0.05$ sample, the peak width is broader
than the resolution-limited case of $x=0.08$ (Fig. 2a). In particular, we
note that the width of the $(0,0,6)$ peak, which is not sensitive to the
in-plane lattice distortion, is resolution limited for all concentrations. \
Hence the peak broadening for the in-plane peaks of the $x=0.05$ sample must
arise from the underlying orthorhombic structure. In addition, we would
expect that the temperature dependence of the $\rm (2,2,0)_{T}$ reflection peak intensity (not integrated intensity) measured by the high-resolution BT-1 should decrease going through
the tetragonal to orthorhombic symmetry change. Figures 2b-d show that this is indeed the case, where 
the tetragonal to orthorhombic symmetry change temperature reduces 
systematically as a function
of increasing F-doping. 

\begin{figure}[tbp]
\includegraphics[scale=1.0]{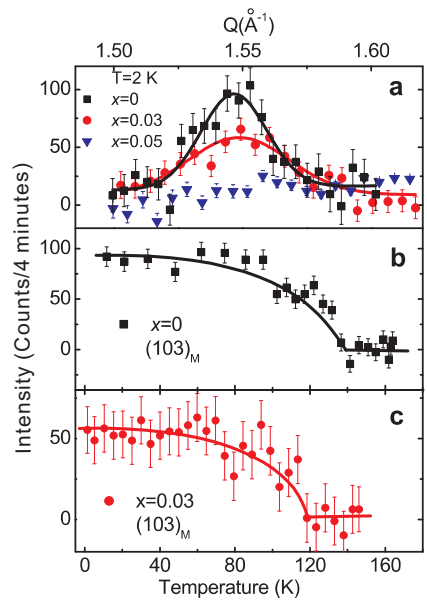}
\caption{(Color online) (a) Wave vector dependence of the AFM ordering peak $
(1,0,3)$ for $x=0,0.03,0.05$ at 2 K. The intensity of scattering is
normalized to the nuclear Bragg peaks and can be compared directly. (b,c)
Temperature dependence of the magnetic scattering for $x=0,0.03$,
respectively. }
\end{figure}

Figure 3 summarizes the F-doping dependence of the AFM Bragg peak and
magnetic order parameter.  $\mu $SR measurements on LaFeAsO$_{1-x}$F$_{x}$
with $x=0,0.03$ \cite{carlo} confirmed that the undoped parent LaFeAsO
compound has static AFM order, but the 3\% F-doping might induce an
incommensurate/stripe-like AFM magnetic order. To determine the F-doping
dependence of the AFM order, we probed the $(1,0,3)$ magnetic peak. Figure
3a plots the wave vector dependence of the $(1,0,3)$ at 2 K. When 3\% F is
introduced, the $(1,0,3)$ peak becomes weaker and broader. The broadening
can be interpreted as a reduction in the Fe spin-spin correlation length
from $208\pm 28$ \AA\ for $x=0$ to $139\pm 33$ \AA\ for $x=0.03$, with the
scattering still being commensurate and centered at $(1,0,3)$ for both
materials. This broadening is somewhat different from the doping-dependent magnetic
scattering for CeFeAsO$_{1-x}$F$_{x}$ \cite{zhao}, where the magnetic peaks
at finite F-dopings were always resolution-limited. This suggests that the
broadening might be interpreted as originating from incommensurate AFM magnetic
order, with an incommensurability that cannot be resolved.  Future experiments on single crystals should
be able to resolve this issue. On further
increasing the F-doping to $x=0.05$, where superconductivity with $T_{c}=8$
K is induced, the $(1,0,3)$ static AFM ordering peak is no longer observable
(Fig. 3a). Therefore, while the orthorhombic lattice distortion extends to
samples with bulk superconductivity, static AFM order does not coexist with
superconductivity within the accuracy of our measurements \cite{takeshita}.

\begin{figure}[tbp]
\includegraphics[scale=0.8]{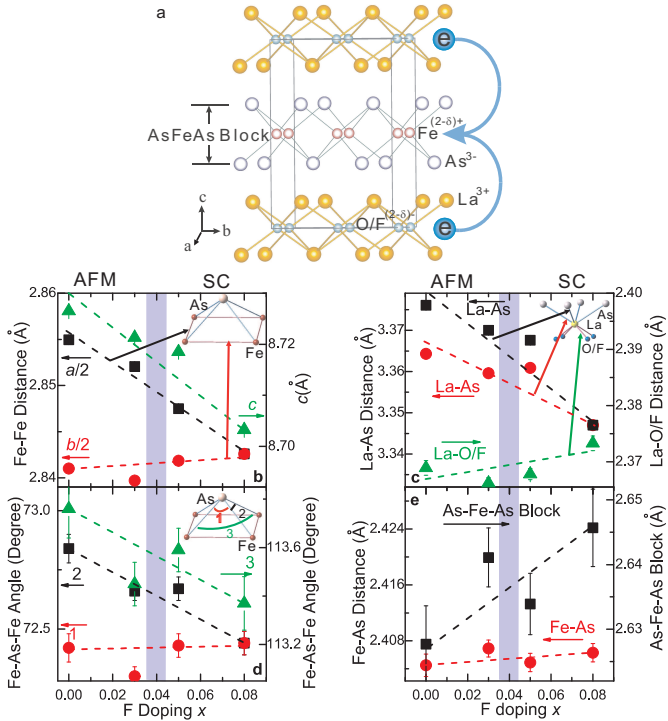}
\caption{(Color online) Low temperature structural evolution of LaFeAsO$
_{1-x}$F$_{x}$ as a function of F doping obtained from analysis of the BT-1
data. There is no sudden structural transition as the AFM order is replaced by the
superconducting phase. The atomic positions of LaFeAsO$_{1-x}$F$_{x}$ and
their temperature dependence are shown in Tables I and II. (a) schematic
diagram defining the As-Fe-As block and illustrating the process of electron
doping. (b) $a$, $b$, $c$ lattice constants of the orthorhombic unit cell
and the two Fe-Fe nearest-neighbor distances as a function of F doping.
Similar to CeFeAsO$_{1-x}$F$_{x}$, F-doping only suppresses the long-axis of
the orthorhombic structure. (c) La-O/F and La-As distances as a function of
F doping. The slight increase in the La-O/F block size is compensated by a
much larger reduction in the La-As distance, resulting in an overall $c$-axis
lattice contraction as shown in (b). d) Fe-As-Fe bond angles as defined in
the inset versus F doping. While angle 1 hardly changes with doping, angles
2 and 3 decrease substantially with increasing F doping. e) The Fe-As bond
distance and As-Fe-As block size versus F doping. The Fe-As distance is
independent of F doping. }
\end{figure}

Figures 3b and 3c show the temperature dependence of the $(1,0,3)$ peak
intensity. Consistent with previous neutron scattering \cite{cruz,mcguire}
and $\mu $SR work \cite{luetkens,carlo}, the N$\mathrm{\acute{e}}$el
temperatures of LaFeAsO$_{1-x}$F$_{x}$ with $x=0,0.03$ are $137\pm 3$ and $%
120\pm 2$ K, respectively. Figure 1c summarizes the structural and magnetic
phase diagram determined from the present work. One of the key differences
between the present phase diagram and that determined by $\mu $SR and M$%
\mathrm{\ddot{o}}$ssbauer effect measurements \cite{luetkens} is the
presence of the orthorhombic lattice distortion in underdoped
superconducting LaFeAsO$_{1-x}$F$_{x}$. This indicates that the evolution
from antiferromagnetism to superconductivity is not directly associated with
the tetragonal to orthorhombic structural phase transition. Instead, our
data appear to support the idea that commensurate AFM order is a competing
ground state to superconductivity, much like the case of electron-doped high-%
$T_{c}$ copper oxides \cite{kang,wilson}. Theoretically, it has been argued
that the orthorhombic lattice distortion in $R$FeAsO$_{1-x}$F$_{x}$ is
associated with nematic ordering of the Fe spin fluctuations and therefore
is a precursor of long range AFM order \cite{fang,xu,xu1}.

\begin{table}[tp]
\caption{Refined crystal structure parameters of LaFeAsO$_{1-x}$F$_x$ with $%
x=0,0.03,0.05$ at 2 K. Space group: $Cmma$. Atomic positions: La: $4g(0,{%
\frac{1}{4}},z)$; Fe: $4b({\frac{1}{4}},0,{\frac{1}{2}})$; As: $4g(0,{\frac{1%
}{4}},z)$; and O/F: $4a({\frac{1}{4}},0,0)$. }
\label{Table}%
\begin{ruledtabular}
\begin{tabular}{cccccccccc}
Atom &   & $x=0$ & $x=0.03$ & $x=0.05$  &\\[3pt]
\hline&  \\[3pt]
  & $a$(\AA) & 0.50988(9) & 5.70407(8) & 5.6995(2)  &\\[3pt]
  & $b$(\AA) & 5.68195(9) & 5.67936(8) & 5.6837(2)  &\\[3pt]
  & $c$(\AA) & 8.7265(1)  & 8.7213(1)  & 8.7185(1)  &\\[3pt]
La& $z$      & 0.1430(3)  & 0.1427(2)  & 0.1431(2)  &\\[3pt]
As& $z$      & 0.6506(3)  & 0.6514(3)  & 0.6510(3)  &\\[3pt]
\hline&  \\[3pt]
 & $R_p$(\%) & 4.26       & 4.36       & 5.21       &\\[3pt]
 & $w_{Rp}$(\%) & 5.47    & 5.75       & 6.87       &\\[3pt]
 & $\chi^2$  & 1.005      & 0.9327     & 1.221      &\\[3pt]
\end{tabular}
\end{ruledtabular}
\end{table}

Previous systematic work on CeFeAsO$_{1-x}$F$_{x}$ \cite{zhao} found that
the impact of F-doping is to compress the $c$- and $a$- axes of the
orthorhombic structure, where $c>a>b$, while leaving the $b$-axis unchanged.
The decrease in the $c$-axis lattice constant is mostly due to the distance
reduction of the CeO and FeAs blocks. To see if this is also true for LaFeAsO$_{1-x}$F$_{x}$, 
we plot the doping dependence of the Fe-Fe distance (Fig.
4b), La-As and La-O/F distances (Fig. 4c), Fe-As-Fe bond angles (Fig. 4d),
and Fe-As/As-Fe-As block distances (Fig. 4e) obtained from detailed analysis
of the high-resolution BT-1 data ( see Tables 1 and 2 for details). Consistent with earlier
work on CeFeAsO$_{1-x}$F$_{x}$ \cite{zhao}, we find that electron doping
suppresses the long $a$-axis of the orthorhombic structure while leaving
the short $b$-axis unchanged. Similarly, doping electrons reduces the
distance between the LaO and FeAs blocks, mostly likely due to increased
Coulomb attraction between these two blocks. Since the Fe-As distance (2.404 
\AA ) is essentially doping independent (Fig. 4e), the net effect of the $a$
-axis lattice contraction is to push the diagonal Fe-As-Fe angle toward the
ideal value of 109.47$^{\circ }$ for the perfect FeAs tetrahedron (Fig. 4d).
The lattice structure is seen to evolve smoothly across the AFM to
superconductivity phase transition. These results confirm the notion that
the most effective way to increase $T_{C}$ in Fe-based superconductors is to
decrease the deviation of the Fe-As(P)-Fe bond angle from the ideal FeAs
tetrahedron \cite{zhao,chlee}.

\section{CONCLUSIONS} 

In summary, we have shown that the orthorhombic lattice distortion present
in undoped LaFeAsO can extend beyond the AFM to superconductivity phase
boundary, whereas the static long-range AFM ordered phase does not seem to
coexist with superconductivity.  The phase diagram of electron-doped LaFeAsO$_{1-x}$F$_{x}$ can therefore be sketched as in Fig. 1c, showing clear coexistence of superconductivity with either the orthorhombic or tetragonal lattice structure.


\section{acknowledgments}

This work is supported by the US NSF through
DMR-0756568, by the US DOE, Division of Materials Science,
Basic Energy Sciences, through DOE DE-FG02-05ER46202. This work is also
supported in part by the US DOE, Division of Scientific
User Facilities, Basic Energy Sciences. The work at the Institute of
Physics, Chinese Academy of Sciences, is supported by the NSF of China, the Chinese Academy of Sciences and the Ministry of
Science and Technology of China.


\end{document}